\documentclass{stacs}

\usepackage{color}

\newcommand{\tinyspace}{\mspace{1mu}}

\newcommand{\bra}[1]{\langle #1|}
\newcommand{\ket}[1]{|#1\rangle}
\newcommand{\braket}[2]{\langle #1|#2\rangle}

\newcommand{\tensor}{\otimes}
\newcommand{\norm}[1]{\left\lVert\tinyspace#1\tinyspace\right\rVert}
\newcommand{\tnorm}[1]{\norm{#1}_{\mathrm{tr}}}
\newcommand{\dnorm}[1]{\norm{#1}_{\diamond}}
\newcommand{\abs}[1]{\left\lvert\tinyspace #1 \tinyspace\right\rvert}

\newcommand{\linear}[1]{\mathbf{L}(\mathcal{#1})}
\newcommand{\density}[1]{\mathbf{D}(\mathcal{#1})}
\newcommand{\unitary}[1]{\mathbf{U}(\mathcal{#1})}

\newcommand{\tr}{\operatorname{tr}}
\newcommand{\F}{\operatorname{F}}
\newcommand{\ptr}[1]{\tr_\mathcal{#1}}

\newcommand{\prb}[1]{\textup{\textsf{#1}}}
\newcommand{\class}[1]{\textup{\textrm{#1}}}

\newenvironment{probenv}[1]{\begin{trivlist}\item {\bf #1.}\em}{\end{trivlist}}

\begin{document}

%% This is to keep the figures from all clumping onto the same page
\setcounter{totalnumber}{1}

%=============================================================================%

\title{Distinguishing Short Quantum Computations}

\author[bill]{Bill Rosgen}{Bill Rosgen}
\address[bill]{Institute for Quantum Computing
         and School of Computer Science \\
         University of Waterloo}
\email{wrosgen@iqc.ca}

\begin{abstract}
  Distinguishing logarithmic depth quantum circuits on mixed states is
  shown to be complete for \class{QIP}, the class of problems having
  quantum interactive proof systems.
  Circuits in this model can represent arbitrary quantum processes,
  and thus this result has implications for the verification of
  implementations of quantum algorithms.
  The distinguishability problem is also complete
  for \class{QIP} on constant depth circuits containing the unbounded
  fan-out gate.
  These results are shown by reducing a \class{QIP}-complete problem to a
  logarithmic depth version of itself using a parallelization technique.
\end{abstract}


\maketitle

%=============================================================================%

\section{Introduction}\label{scn:intro}

Much of the difficulty in implementing quantum algorithms in practice is that
qubits quickly decohere upon interacting with the environment.  This
entanglement destroying process limits the length of the
computations that can be realized by experiment.
Implementing quantum algorithms as circuits of low
depth can provide a way to perform as much computation as possible
within the limited time available, and  
for this reason there is significant interest in finding
short quantum circuits for important problems.

Log-depth quantum circuits have been found for several significant problems
including the approximate quantum Fourier transform~\cite{CleveW00}
and the encoding and decoding operations for many quantum error
correcting codes~\cite{MooreN02}.
In addition to these applications, a procedure for parallelizing to log-depth
a large class of quantum circuits has recently been 
discovered~\cite{BroadbentK07}.  These examples demonstrate the
surprising power of short quantum circuits.

Much of the work on quantum circuits is done in the standard model 
of unitary quantum circuits on pure states.
In this paper a slightly different model of computation is
considered:  the model of mixed state quantum circuits,
introduced by Aharonov, Kitaev, and Nisan~\cite{AharonovK+98}.  
While much of the previous complexity-theoretic work on short quantum
circuits has been in the unitary 
model~\cite{FennerG+05, GreenH+02},
there has also been work outside of this model~\cite{TehralD04}.
There are several advantages to
considering the more general model of mixed state circuits.
The primary advantage is that the mixed state model is able to capture
any process allowed by quantum mechanics, so that
results on this model may have implications for experimental work in
quantum computing.
The problem of distinguishing circuits may thus be thought of as the
problem of distinguishing potentially noisy physical processes.
As an example, finding an error in an implementation of a quantum 
algorithm is simply the problem of distinguishing the
constructed circuit from one that is known to be correct.

Unfortunately, in this paper it is shown that the apparent power of
short quantum computations comes with a price:
logarithmic depth quantum circuits are \emph{exactly} as difficult
to distinguish as polynomial depth quantum circuits.
This equivalence implies the surprising result that distinguishing
log-depth quantum computations is complete for the class \class{QIP},
the set of all problems that have quantum interactive
proof systems.
As
$\class{PSPACE} \subseteq \class{QIP}
\subseteq \class{EXP}$~\cite{KitaevW00}, this result also 
implies that the problem is \class{PSPACE}-hard.

The result on circuit distinguishability 
is shown using the closely related problem of determining
if two circuits can be made to output states that are close together.
This problem was introduced by Kitaev and Watrous~\cite{KitaevW00} 
who show it to
be both complete for \class{QIP} and contained in \class{EXP}.  
The main result of the present paper is obtained by reducing an 
instance of this problem of polynomial depth to an equivalent 
instance of logarithmic depth.
This demonstrates that the problem of close images remains complete for
\class{QIP} even under a logarithmic depth restriction.
The hardness of distinguishing short quantum circuits is then
demonstrated by a modification to the argument in~\cite{RosgenW05} to
show that the equivalence of close images problem and the
distinguishability holds even for log-depth circuits.

The remainder of this paper is organized as follows.  In the next
section, some of the notation and results that will be needed are summarized.
This is followed by Section~\ref{scn:problems}, where the
complete problems for \class{QIP} are discussed.  In
Section~\ref{scn:construction} the reduction from the polynomial depth
to logarithmic depth versions of the close images problem is given,
and the correctness of this construction is shown in
Section~\ref{scn:soundness}.  The equivalence between the log-depth
close images problem and the problem of distinguishing log-depth
computations is discussed in Section~\ref{scn:distinguish}.

%=============================================================================%

\section{Preliminaries}\label{scn:prelim}

This section outlines some of the definitions and results that
will be used throughout the paper.  For a more thorough treatment of
the concepts introduced here see~\cite{KitaevS+02} and~\cite{NielsenC00}.

Throughout the paper scripted letters such as $\mathcal{H}$ will refer
to finite dimensional Hilbert spaces, $\density{H}$ will denote
the set of all density matrices on $\mathcal{H}$, and $\unitary{H,K}$
will denote the norm-preserving linear operators from $\mathcal{H}$ to
$\mathcal{K}$.
The proof of the main result will make extensive use of 
two notions of distance between quantum
states.  The first of these is the fidelity.  The \emph{fidelity} between two
positive semidefinite operators $X$ and $Y$ on a space $\mathcal{H}$ 
can be defined as
%\[ \F(X,Y) = \tr \sqrt{\sqrt{X} Y \sqrt{X}}. \]
%An alternate characterization of the fidelity, known as Uhlmann's
%Theorem, will be useful later.  This is given by
\begin{equation*}
  F(X,Y) = \max \{ \abs{ \braket{\phi}{\psi} } : 
  \ket\phi, \ket\psi \in \mathcal{H \tensor K}, 
  \ptr{K} \ket\phi\bra\phi = X, 
  \ptr{K} \ket\psi\bra\psi = Y\}.
\end{equation*}
This definition is known as Uhlmann's Theorem, and it is used here as
it is more directly applicable to the task at hand than the usual definition.
As any purification of a state necessarily purifies the partial
trace of that state, this equation implies that the fidelity is
nondecreasing under the partial trace.  
This property is known as \emph{monotonicity}
and can be stated more formally as
$ F(X,Y) \leq F(\ptr{K} X, \ptr{K} Y) $
where $X,Y$ are positive semidefinite operators on $\mathcal{H \tensor K}$.
The final property of the fidelity that will be needed is the
result that the maximum fidelity of
any outputs of two transformations is multiplicative with respect to
the tensor product.  This result can be found in~\cite{KitaevS+02} 
(see Problem~11.10 and apply the multiplicativity of the diamond
norm with respect to the tensor product).
\begin{theorem}[Kitaev, Shen, and Vyalyi~\cite{KitaevS+02}]
\label{thm:outfid-mult}
    For any completely positive transformations $\Phi_1, \Phi_2,
    \Psi_1, \Psi_2$ on states in $\mathcal{H}$
    \[ \max_{\rho, \xi \in \density{H \tensor H}} 
       \F \left( (\Phi_1 \tensor  \Phi_2) (\rho), 
           (\Psi_1 \tensor \Psi_2) (\xi)\right)
       = \prod_{i=1,2} \max_{\rho, \xi \in \density{H}}
           \F(\Phi_i (\rho), \Psi_i (\xi)) \]    
\end{theorem}

The second notion of distance that will be used is the \emph{trace norm},
which can be defined for any linear operator $X$ by 
$ \tnorm{X} = \tr \sqrt{X^* X}, $
or equivalently as the sum of the singular
values of $X$.  This quantity is a norm, and so in particular it
satisfies the triangle inequality.  Similar to the fidelity, the trace
norm is monotone under the partial trace, though in this case the
trace norm is non-increasing under this operation.
The proofs that follow will make essential
use of the Fuchs-van de Graaf Inequalities~\cite{FuchsG99}
that relate the trace norm and the fidelity.  
For any density operators $\rho$ and $\xi$ on the
same space, these inequalities are
%% \begin{equation*}
%%   1 - \frac{1}{2}\tnorm{\rho - \xi}
%%   \leq \F(\rho,\xi)
%%   \leq \sqrt{ 1 - \frac{1}{4}\tnorm{\rho - \xi}^2},
%% \end{equation*}
%% or equivalently
\begin{equation*}
  1 - \F(\rho,\xi)
  \leq \frac{1}{2} \tnorm{\rho - \xi}
  \leq \sqrt{1 - \F(\rho,\xi)}.
\end{equation*}

In addition to these measures on quantum states, it will be helpful to
have a distance measure on quantum transformations.  One such measure
is the \emph{diamond norm}, which for a completely positive transformation
$\Phi$ on density operators on $\mathcal{H}$ is given by
\[ \dnorm{\Phi} = \sup_{\rho \in \density{H \tensor H}} \tnorm{(\Phi \tensor
   I_{\linear{H}})(\rho)}. \]
This norm is essential when considering transformations as it
represents the distinguishability of two transformations when a
reference system is taken into account.  
The simple supremum of the
trace norm over all inputs to the channel is not stable under
the addition of a reference system, and so the diamond norm is used 
in place of the simpler one.  
More properties and a more thorough definition of this
norm can be found in~\cite{KitaevS+02}.

The circuit model that will be used in this paper is the mixed state model
introduced by Aharonov, Kitaev, and Nisan~\cite{AharonovK+98}.
Circuits in this model are composed of qubits that are acted upon by
arbitrary trace preserving and completely positive operations.  This
model allows for non-unitary operations, such as measurement or the
introduction of ancillary qubits, to occur in the middle of the
circuit.
It is important to note that this model captures any physical process
that quantum mechanics allows, and so in particular, any computation
that can be done on mixed states with measurements can be represented
in this model.
Fortunately this model is polynomially equivalent to the standard
model of unitary quantum circuits (with ancilla) followed by
measurement, as shown in~\cite{AharonovK+98}.  
This will allow us to consider only circuits composed of unitary gates
from some finite basis of one and two qubit gates with the additional
operations of introducing qubits in the $\ket 0$ state and measuring
in the computational basis.  
This restriction can be strengthened,
again with no loss of generality, to assume that all ancillary qubits
are introduced at the start of the circuit and that all measurements
are performed at the end.

We will often add to this circuit model one additional gate:
the unbounded fan-out gate.
This gate, in constant depth, applies a controlled-not
operation from one qubit to an arbitrary number of output qubits.
It is not clear that this gate is a reasonable choice in a
standard basis of gates, and so it will be clearly marked when this gate
is allowed into the circuit model under consideration.
As an example of the power
of this gate it can be used to build a constant depth circuit for the
approximate quantum Fourier transform~\cite{HoyerS05}.
This gate is considered here for the sole reason that if it is
included in the standard set of gates, the main result will also
hold for constant depth circuits.

\begin{figure}
  \begin{center}
\setlength{\unitlength}{3947sp}%
\begingroup\makeatletter\ifx\SetFigFont\undefined%
\gdef\SetFigFont#1#2#3#4#5{%
  \reset@font\fontsize{#1}{#2pt}%
  \fontfamily{#3}\fontseries{#4}\fontshape{#5}%
  \selectfont}%
\fi\endgroup%
%
% This has been hacked so that the figure is centered on the swap gate
% if the figure is spilling into the right margin, this is the problem
% origninal bounding box:
%\begin{picture}(3810,1899)(1339,-1423)
%
\begin{picture}(3300,1899)(1339,-1423)
\put(4426,164){\makebox(0,0)[lb]{\smash{{\SetFigFont{12}{14.4}{\familydefault}{\mddefault}{\updefault}{\color[rgb]{0,0,0}Measured}%
}}}}
\thinlines
{\color[rgb]{0,0,0}\put(1351,-361){\line( 1, 0){1350}}
}%
{\color[rgb]{0,0,0}\put(1351,-511){\line( 1, 0){1350}}
}%
{\color[rgb]{0,0,0}\put(1351,-586){\line( 1, 0){1350}}
}%
{\color[rgb]{0,0,0}\put(1351,-661){\line( 1, 0){1350}}
}%
{\color[rgb]{0,0,0}\put(1351,-736){\line( 1, 0){1350}}
}%
{\color[rgb]{0,0,0}\put(1351,-961){\line( 1, 0){1350}}
}%
{\color[rgb]{0,0,0}\put(1351,-886){\line( 1, 0){1350}}
}%
{\color[rgb]{0,0,0}\put(1351,-1036){\line( 1, 0){1350}}
}%
{\color[rgb]{0,0,0}\put(1351,-1111){\line( 1, 0){1350}}
}%
{\color[rgb]{0,0,0}\put(1351,-1186){\line( 1, 0){1350}}
}%
{\color[rgb]{0,0,0}\put(1351,-1261){\line( 1, 0){1350}}
}%
{\color[rgb]{0,0,0}\put(3301,-961){\line( 1, 0){1350}}
}%
{\color[rgb]{0,0,0}\put(3301,-886){\line( 1, 0){1350}}
}%
{\color[rgb]{0,0,0}\put(3301,-1036){\line( 1, 0){1350}}
}%
{\color[rgb]{0,0,0}\put(3301,-1111){\line( 1, 0){1350}}
}%
{\color[rgb]{0,0,0}\put(3301,-1186){\line( 1, 0){1350}}
}%
{\color[rgb]{0,0,0}\put(3301,-1261){\line( 1, 0){1350}}
}%
{\color[rgb]{0,0,0}\put(3301,-436){\line( 1, 0){1350}}
}%
{\color[rgb]{0,0,0}\put(3301,-361){\line( 1, 0){1350}}
}%
{\color[rgb]{0,0,0}\put(3301,-511){\line( 1, 0){1350}}
}%
{\color[rgb]{0,0,0}\put(3301,-586){\line( 1, 0){1350}}
}%
{\color[rgb]{0,0,0}\put(3301,-661){\line( 1, 0){1350}}
}%
{\color[rgb]{0,0,0}\put(3301,-736){\line( 1, 0){1350}}
}%
{\color[rgb]{0,0,0}\put(3001,239){\circle*{76}}
}%
{\color[rgb]{0,0,0}\put(2701,-1411){\framebox(600,1200){$W$}}
}%
{\color[rgb]{0,0,0}\put(3601,239){\line(-1, 0){1200}}
}%
{\color[rgb]{0,0,0}\put(3001,239){\line( 0,-1){450}}
}%
{\color[rgb]{0,0,0}\put(1951, 14){\framebox(450,450){$H$}}
}%
{\color[rgb]{0,0,0}\put(3601, 14){\framebox(450,450){$H$}}
}%
{\color[rgb]{0,0,0}\put(4051,239){\line( 1, 0){300}}
}%
{\color[rgb]{0,0,0}\put(1651,239){\line( 1, 0){300}}
}%
\put(1351,164){\makebox(0,0)[lb]{\smash{{\SetFigFont{12}{14.4}{\familydefault}{\mddefault}{\updefault}{\color[rgb]{0,0,0}$\ket 0$}%
}}}}
{\color[rgb]{0,0,0}\put(1351,-436){\line( 1, 0){1350}}
}%
\end{picture}%
  \end{center}
  \caption{A circuit implementing the swap test.}
  \label{fig:swaptest}
\end{figure}
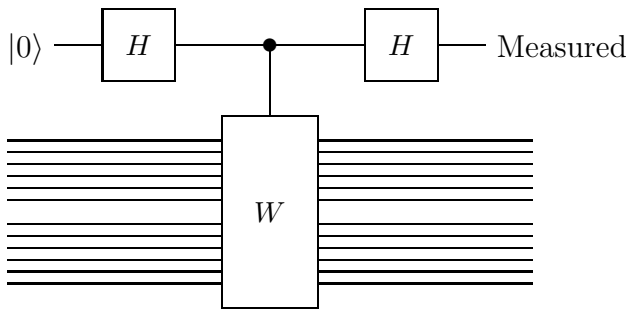

For spaces $\mathcal{H}$ and $\mathcal{K}$ of the same dimension, we
use $W \in \unitary{H \tensor K, \mathcal{H \tensor K}}$ to represent 
the operation that swaps the states in the two spaces.  
As $W$ is a permutation matrix when
expressed in the computational basis, and the permutation that it
encodes is composed exclusively of transpositions, the swap operation is
both hermitian and unitary.  Furthermore, $W$ can easily be
implemented in constant depth, as all of the required transpositions can
be performed in parallel.
This operator is the essential
component of the \emph{swap test}, where a controlled $W$ operation is
used to determine how close two states are to each other.
A circuit performing the swap test is given in
Figure~\ref{fig:swaptest}, where the measurement is performed in the
computational basis.
Another way to view the swap test is as a projective measurement
onto the symmetric and antisymmetric subspaces.
The projections in this measurement are
given by $(I + W)/2$ and $(I-W)/2$.  
This formulation of the swap test is equivalent to the circuit presented in
Figure~\ref{fig:swaptest}.

It is not immediately clear how a controlled operation on $n$ qubits,
such as the controlled-swap operation used in the swap test, 
can be performed in depth logarithmic in $n$.
The straightforward implementation requires using one control qubit to
control each of the gates in the operation.
However, Moore and
Nilsson~\cite{MooreN02} give a simple construction that allows such an
operation to be performed in log-depth.
\begin{prop}[Moore and Nilsson]
  Any log depth operation on $n$ qubits controlled by one qubit
  can be implemented in $O(\log n)$ depth with $O(n)$ ancillary qubits.
\end{prop}
Moore and Nilsson prove this only for the constant depth case, but the
method of proof that they use immediately extends to the log depth
case.  They prove this proposition by using a tree of $\log n$
controlled-not operations to `duplicate' the control qubit.  These
copies can then be used to control the remaining operations, with each
control qubit used at most a logarithmic number of times.
This proposition, as an example, implies that the
swap test circuit on $n$ qubits shown in Figure~\ref{fig:swaptest} can
be implemented in depth $O(\log n)$.

If the fan-out gate is allowed into the standard basis of gates, then
controlled operations can be performed with only constant depth
overhead.
A circuit that performs this can be obtained by simply using one
fan-out gate to make $n$ copies (in the computational basis) of the
control qubit onto ancillary qubits.
These `copies' may then be used to control each of the $n$ operations,
with a final application of the fan-out gate to restore the ancillary
qubits to the $\ket 0$ state.
As controlled operations will be the only place that the circuits
constructed here exceed constant depth, this will allow the proof of
the main result for constant depth circuits with fan-out.

%=============================================================================%

\section{Complete Problems for \class{QIP}}\label{scn:problems}

The \prb{Close Images} problem, defined and shown to be complete for
\class{QIP} in~\cite{KitaevW00} can be stated as follows.
\begin{probenv}{Close Images}
  For constants $0 < b < a \leq 1$, the input consists of quantum
  circuits $Q_1$ and $Q_2$ that implement transformations
  from $\mathcal{H}$ to $\mathcal{K}$.
  The promise problem is to distinguish the two cases:
  \begin{description}
    \item[Yes] $\F(Q_1(\rho), Q_2(\xi)) \geq a$ for some $\rho, \xi
      \in \density{H}$,
    \item[No] $\F(Q_1(\rho), Q_2(\xi)) \leq b$ for all $\rho, \xi
    \in \density{H}$.
  \end{description}
\end{probenv}
This is simply the problem of determining if there are inputs to $Q_1$
and $Q_2$ that cause them to output states that are nearly the same.
It will be helpful to abbreviate the name of this problem as $\prb{CI}_{a,b}$.

A closely related problem is that of distinguishing two quantum
circuits.  This problem was introduced and shown complete for
\class{QIP} in~\cite{RosgenW05}.
\begin{probenv}{Quantum Circuit Distinguishability}
  For constants $0 \leq b < a \leq 2$, the input consists of quantum
  circuits $Q_1$ and $Q_2$ that implement transformations from
  $\mathcal{H}$ to $\mathcal{K}$.
  The promise problem is to distinguish the two cases:
  \begin{description}
    \item[Yes] $\dnorm{Q_1 - Q_2} \geq a$,
    \item[No] $\dnorm{Q_1 - Q_2} \leq b$.
  \end{description}
\end{probenv}
Less formally, this problem asks: is there an input density matrix
$\rho$ on which the circuits $Q_1$ and $Q_2$ can be made to act differently?
This problem will be referred to as $\prb{QCD}_{a,b}$.

It is our aim to prove that these problems remain complete for
\class{QIP} when restricted to circuits $Q_1$ and $Q_2$ that are of
depth logarithmic in the number of input qubits.  This will be
achieved in the case of perfect soundness error, i.e. $a = 1,2$
in the above problem definitions.  Both of these problem remain
complete for \class{QIP} in this case.
This restriction serves only to make these problems easier, as
distinguishing the two cases for a weaker promise can only be more
difficult, so 
the results of this paper will also imply the hardness of
the more general problems.
The log-depth versions of
these problems will be referred to as 
$\prb{Log-depth CI}_{1,b}$ and $\prb{Log-depth QCD}_{2,b}$, 
and since these are restrictions of
\class{QIP}-complete problems it is clear that they are also in
\class{QIP}.
Similarly, the abbreviations $\prb{Const-depth CI}_{1,b}$ and
$\prb{Const-depth QCD}_{2,b}$ for the
versions of these problems on constant-depth circuits will be
convenient.

%=============================================================================%

\section{Log-Depth Construction}\label{scn:construction}

In this section the reduction from the general
$\prb{CI}_{1,b}$ problem to the log-depth restriction of the
problem is described.  
The general idea behind the construction is to simply slice
the circuits of an instance of $\prb{CI}_{1,b}$ into
logarithmic-depth pieces and run them in parallel.  These circuits
will require more input, but if each piece of the circuit is given as
input the same state output by the previous piece, then the output of the
last piece of the circuit will be equal to the output of the original
circuit.  
This may not be
the case if the intermediate inputs are not the outputs of the
previous pieces, and so additional tests that ensure
these inputs are at least close to the desired states are required.

To describe the reduction, let $Q_1$ and $Q_2$ be the circuits from
an instance of $\prb{CI}_{1,b}$, and let $n$ be the size (number of
gates) of $Q_1$ and $Q_2$ (by padding the smaller circuit, if necessary).
In order to perform the slicing of the circuit into pieces it is assumed
that $Q_1$ and $Q_2$ first introduce any necessary
ancillary qubits, then apply local unitary gates, and finally trace
out any qubits that are not part of the input.
This restriction can
be made with no loss in generality, as any quantum circuit, even one
that incorporates measurements and other non-unitary operations, can
be approximated by such a circuit, and furthermore, this circuit uses
a number of gates that is a polynomial in the size of the original
circuit~\cite{AharonovK+98}.

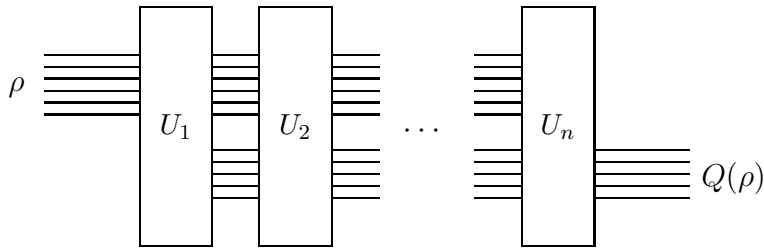
\begin{figure}
  \begin{center}
\setlength{\unitlength}{3947sp}%
\begingroup\makeatletter\ifx\SetFigFont\undefined%
\gdef\SetFigFont#1#2#3#4#5{%
  \reset@font\fontsize{#1}{#2pt}%
  \fontfamily{#3}\fontseries{#4}\fontshape{#5}%
  \selectfont}%
\fi\endgroup%
\begin{picture}(4350,1524)(376,-1573)
\put(4726,-1186){\makebox(0,0)[lb]{\smash{{\SetFigFont{12}{14.4}{\familydefault}{\mddefault}{\updefault}{\color[rgb]{0,0,0}$Q(\rho)$}%
}}}}
\thinlines
{\color[rgb]{0,0,0}\put(1951,-1561){\framebox(450,1500){$U_2$}}
}%
{\color[rgb]{0,0,0}\put(3601,-1561){\framebox(450,1500){$U_n$}}
}%
{\color[rgb]{0,0,0}\put(901,-361){\line( 1, 0){300}}
}%
{\color[rgb]{0,0,0}\put(901,-436){\line( 1, 0){300}}
}%
{\color[rgb]{0,0,0}\put(901,-511){\line( 1, 0){300}}
}%
{\color[rgb]{0,0,0}\put(901,-586){\line( 1, 0){300}}
}%
{\color[rgb]{0,0,0}\put(901,-661){\line( 1, 0){300}}
}%
{\color[rgb]{0,0,0}\put(901,-736){\line( 1, 0){300}}
}%
{\color[rgb]{0,0,0}\put(1651,-361){\line( 1, 0){300}}
}%
{\color[rgb]{0,0,0}\put(1651,-436){\line( 1, 0){300}}
}%
{\color[rgb]{0,0,0}\put(1651,-511){\line( 1, 0){300}}
}%
{\color[rgb]{0,0,0}\put(1651,-586){\line( 1, 0){300}}
}%
{\color[rgb]{0,0,0}\put(1651,-661){\line( 1, 0){300}}
}%
{\color[rgb]{0,0,0}\put(1651,-736){\line( 1, 0){300}}
}%
{\color[rgb]{0,0,0}\put(1651,-961){\line( 1, 0){300}}
}%
{\color[rgb]{0,0,0}\put(1651,-1036){\line( 1, 0){300}}
}%
{\color[rgb]{0,0,0}\put(1651,-1111){\line( 1, 0){300}}
}%
{\color[rgb]{0,0,0}\put(1651,-1186){\line( 1, 0){300}}
}%
{\color[rgb]{0,0,0}\put(1651,-1261){\line( 1, 0){300}}
}%
{\color[rgb]{0,0,0}\put(2401,-361){\line( 1, 0){300}}
}%
{\color[rgb]{0,0,0}\put(2401,-436){\line( 1, 0){300}}
}%
{\color[rgb]{0,0,0}\put(2401,-511){\line( 1, 0){300}}
}%
{\color[rgb]{0,0,0}\put(2401,-586){\line( 1, 0){300}}
}%
{\color[rgb]{0,0,0}\put(2401,-661){\line( 1, 0){300}}
}%
{\color[rgb]{0,0,0}\put(2401,-736){\line( 1, 0){300}}
}%
{\color[rgb]{0,0,0}\put(2401,-961){\line( 1, 0){300}}
}%
{\color[rgb]{0,0,0}\put(2401,-1036){\line( 1, 0){300}}
}%
{\color[rgb]{0,0,0}\put(2401,-1111){\line( 1, 0){300}}
}%
{\color[rgb]{0,0,0}\put(2401,-1186){\line( 1, 0){300}}
}%
{\color[rgb]{0,0,0}\put(2401,-1261){\line( 1, 0){300}}
}%
{\color[rgb]{0,0,0}\put(3301,-361){\line( 1, 0){300}}
}%
{\color[rgb]{0,0,0}\put(3301,-436){\line( 1, 0){300}}
}%
{\color[rgb]{0,0,0}\put(3301,-511){\line( 1, 0){300}}
}%
{\color[rgb]{0,0,0}\put(3301,-586){\line( 1, 0){300}}
}%
{\color[rgb]{0,0,0}\put(3301,-661){\line( 1, 0){300}}
}%
{\color[rgb]{0,0,0}\put(3301,-736){\line( 1, 0){300}}
}%
{\color[rgb]{0,0,0}\put(3301,-961){\line( 1, 0){300}}
}%
{\color[rgb]{0,0,0}\put(3301,-1036){\line( 1, 0){300}}
}%
{\color[rgb]{0,0,0}\put(3301,-1111){\line( 1, 0){300}}
}%
{\color[rgb]{0,0,0}\put(3301,-1186){\line( 1, 0){300}}
}%
{\color[rgb]{0,0,0}\put(3301,-1261){\line( 1, 0){300}}
}%
{\color[rgb]{0,0,0}\put(4351,-1036){\line( 1, 0){300}}
}%
{\color[rgb]{0,0,0}\put(4351,-961){\line( 1, 0){300}}
}%
{\color[rgb]{0,0,0}\put(4351,-1111){\line( 1, 0){300}}
}%
{\color[rgb]{0,0,0}\put(4351,-1186){\line( 1, 0){300}}
}%
{\color[rgb]{0,0,0}\put(4351,-1261){\line( 1, 0){300}}
}%
{\color[rgb]{0,0,0}\put(4051,-961){\line( 1, 0){300}}
}%
{\color[rgb]{0,0,0}\put(4051,-1036){\line( 1, 0){300}}
}%
{\color[rgb]{0,0,0}\put(4051,-1111){\line( 1, 0){300}}
}%
{\color[rgb]{0,0,0}\put(4051,-1186){\line( 1, 0){300}}
}%
{\color[rgb]{0,0,0}\put(4051,-1261){\line( 1, 0){300}}
}%
{\color[rgb]{0,0,0}\put(601,-361){\line( 1, 0){300}}
}%
{\color[rgb]{0,0,0}\put(601,-511){\line( 1, 0){300}}
}%
{\color[rgb]{0,0,0}\put(601,-436){\line( 1, 0){300}}
}%
{\color[rgb]{0,0,0}\put(601,-586){\line( 1, 0){300}}
}%
{\color[rgb]{0,0,0}\put(601,-661){\line( 1, 0){300}}
}%
{\color[rgb]{0,0,0}\put(601,-736){\line( 1, 0){300}}
}%
\put(376,-586){\makebox(0,0)[lb]{\smash{{\SetFigFont{12}{14.4}{\familydefault}{\mddefault}{\updefault}{\color[rgb]{0,0,0}$\rho$}%
}}}}
\put(2851,-886){\makebox(0,0)[lb]{\smash{{\SetFigFont{12}{14.4}{\familydefault}{\mddefault}{\updefault}{\color[rgb]{0,0,0}$\cdots$}%
}}}}
{\color[rgb]{0,0,0}\put(1201,-1561){\framebox(450,1500){$U_1$}}
}%
\end{picture}%
  \end{center}
  \caption{The original circuits $Q_1$ and $Q_2$ decomposed
    into constant depth unitary circuits.}\label{fig:original}
\end{figure}

A simple way to decompose $Q_1$ into constant depth pieces is to
simply let each gate of $Q_1$ be a piece in the decomposition.
Let $U_1, U_2, \ldots, U_n$ be these
pieces, with the additional complication that the operation $U_1$ both adds the
ancillary qubits and performs the first gate of the circuit.
In a similar way, $Q_2$ can be decomposed into constant depth pieces $V_1,
V_2, \ldots, V_n$.
These pieces are shown in Figure~\ref{fig:original}.
If $Q_1$ and $Q_2$ implement transformations from $\mathcal{H}$
to $\mathcal{K}$, using ancillary qubits that fit into $\mathcal{A}$,
and trace out the qubits in $\mathcal{B}$, then the spaces $\mathcal{H
  \tensor A}$ and $\mathcal{B \tensor K}$ are isomorphic, since by assumption
$Q_1$ and $Q_2$ first introduce any needed ancilla
and only trace qubits out at the end of the computation.
Using these spaces, and implicitly this isomorphism, we have
\begin{align*}
  U_1, V_1 & \in \mathbf{U}(
      \mathcal{H}_1, \mathcal{B}_1 \tensor \mathcal{K}_1) \\
  U_i, V_i & \in \mathbf{U}(
      \mathcal{H}_i \tensor \mathcal{A}_i, \mathcal{B}_i \tensor \mathcal{K}_i)
      \quad \text{for $2 \leq i \leq n$},
\end{align*}
where the subscripted spaces are copies of the
non-subscripted spaces that hold the input or output of one of the
pieces of the original circuits.
As an example of this notation, if $\rho \in \density{H}$, then the
output of $Q_1$ on $\rho$ is given by
\[ \tr_{\mathcal{B}_n} U_n U_{n-1} \cdots U_1 \rho U_1^* U_2^* \cdots U_n^*, \]
and the output of $Q_2$ is given by the same expression using the
$V_i$ operators.

\begin{figure}
  \begin{center}
\setlength{\unitlength}{3947sp}%
\begingroup\makeatletter\ifx\SetFigFont\undefined%
\gdef\SetFigFont#1#2#3#4#5{%
  \reset@font\fontsize{#1}{#2pt}%
  \fontfamily{#3}\fontseries{#4}\fontshape{#5}%
  \selectfont}%
\fi\endgroup%
\begin{picture}(4200,2049)(-524,-2623)
\put(-374,-1486){\makebox(0,0)[lb]{\smash{{\SetFigFont{12}{14.4}{\familydefault}{\mddefault}{\updefault}{\color[rgb]{0,0,0}$\ket{\psi_i}$}%
}}}}
\thinlines
{\color[rgb]{0,0,0}\put(601,-1861){\line( 1, 0){600}}
}%
{\color[rgb]{0,0,0}\put(601,-1936){\line( 1, 0){600}}
}%
{\color[rgb]{0,0,0}\put(601,-1786){\line(-1, 0){600}}
}%
{\color[rgb]{0,0,0}\put(601,-1861){\line(-1, 0){600}}
}%
{\color[rgb]{0,0,0}\put(601,-1936){\line(-1, 0){600}}
}%
{\color[rgb]{0,0,0}\put(601,-2311){\line( 1, 0){600}}
}%
{\color[rgb]{0,0,0}\put(601,-2386){\line( 1, 0){600}}
}%
{\color[rgb]{0,0,0}\put(601,-2461){\line( 1, 0){600}}
}%
{\color[rgb]{0,0,0}\put(601,-2311){\line(-1, 0){600}}
}%
{\color[rgb]{0,0,0}\put(601,-2386){\line(-1, 0){600}}
}%
{\color[rgb]{0,0,0}\put(601,-2461){\line(-1, 0){600}}
}%
{\color[rgb]{0,0,0}\put(2626,-811){\circle*{76}}
}%
{\color[rgb]{0,0,0}\put(1426,-811){\circle*{76}}
}%
{\color[rgb]{0,0,0}\put(1651,-1336){\line( 1, 0){150}}
}%
{\color[rgb]{0,0,0}\put(1651,-1411){\line( 1, 0){150}}
}%
{\color[rgb]{0,0,0}\put(1651,-1486){\line( 1, 0){150}}
}%
{\color[rgb]{0,0,0}\put(1801,-1636){\framebox(450,450){$U_i$}}
}%
{\color[rgb]{0,0,0}\put(2401,-2011){\framebox(450,825){\parbox{0.6in}{\scriptsize\centering{swap\\[-1mm] test}}}}
}%
{\color[rgb]{0,0,0}\put(2851,-1636){\line( 1, 0){600}}
}%
{\color[rgb]{0,0,0}\put(3451,-1636){\line( 1, 0){150}}
}%
{\color[rgb]{0,0,0}\put(1651,-1786){\line( 1, 0){750}}
}%
{\color[rgb]{0,0,0}\put(1651,-1861){\line( 1, 0){750}}
}%
{\color[rgb]{0,0,0}\put(1651,-1936){\line( 1, 0){750}}
}%
{\color[rgb]{0,0,0}\put(2251,-1336){\line( 1, 0){150}}
}%
{\color[rgb]{0,0,0}\put(2251,-1411){\line( 1, 0){150}}
}%
{\color[rgb]{0,0,0}\put(2251,-1486){\line( 1, 0){150}}
}%
{\color[rgb]{0,0,0}\put(1651,-2086){\line( 1, 0){1950}}
}%
{\color[rgb]{0,0,0}\put(1801,-2611){\framebox(450,450){$U_{i+1}$}}
}%
{\color[rgb]{0,0,0}\put(1651,-2311){\line( 1, 0){150}}
}%
{\color[rgb]{0,0,0}\put(1651,-2386){\line( 1, 0){150}}
}%
{\color[rgb]{0,0,0}\put(1651,-2461){\line( 1, 0){150}}
}%
{\color[rgb]{0,0,0}\put(2251,-2311){\line( 1, 0){150}}
}%
{\color[rgb]{0,0,0}\put(2251,-2386){\line( 1, 0){150}}
}%
{\color[rgb]{0,0,0}\put(2251,-2461){\line( 1, 0){150}}
}%
{\color[rgb]{0,0,0}\put(1201,-2536){\framebox(450,825){\parbox{0.6in}{\scriptsize\centering{swap\\[-1mm] test}}}}
}%
{\color[rgb]{0,0,0}\put(1651,-1336){\line(-1, 0){1650}}
}%
{\color[rgb]{0,0,0}\put(1651,-1411){\line(-1, 0){1650}}
}%
{\color[rgb]{0,0,0}\put(  1,-1486){\line( 1, 0){1650}}
}%
{\color[rgb]{0,0,0}\put(2401,-2311){\line( 1, 0){1200}}
}%
{\color[rgb]{0,0,0}\put(2401,-2386){\line( 1, 0){1200}}
}%
{\color[rgb]{0,0,0}\put(2401,-2461){\line( 1, 0){1200}}
}%
%{\color[rgb]{0,0,0}\put(601,-811){\line( 1, 0){2025}}
%}%
{\color[rgb]{0,0,0}\put(2626,-1186){\line( 0, 1){375}}
}%
{\color[rgb]{0,0,0}\put(1426,-1711){\line( 0, 1){900}}
}%
{\color[rgb]{0,0,0}\put(601,-1786){\line( 1, 0){600}}
}%
{\color[rgb]{0,0,0}\put(1801,-1036){\framebox(450,450){$X$}}
}%
{\color[rgb]{0,0,0}\put(601,-1036){\framebox(450,450){$H$}}
}%
{\color[rgb]{0,0,0}\put(451,-811){\line( 1, 0){150}}
}%
{\color[rgb]{0,0,0}\put(1051,-811){\line( 1, 0){750}}
}%
{\color[rgb]{0,0,0}\put(2251,-811){\line( 1, 0){375}}
}%
\put(151,-886){\makebox(0,0)[lb]{\smash{{\SetFigFont{12}{14.4}{\familydefault}{\mddefault}{\updefault}{\color[rgb]{0,0,0}$\ket 0$}%
}}}}
\put(3676,-2461){\makebox(0,0)[lb]{\smash{{\SetFigFont{12}{14.4}{\familydefault}{\mddefault}{\updefault}{\color[rgb]{0,0,0}$\ket{\psi_{i+2}}$}%
}}}}
\put(-524,-1936){\makebox(0,0)[lb]{\smash{{\SetFigFont{12}{14.4}{\familydefault}{\mddefault}{\updefault}{\color[rgb]{0,0,0}$\ket{\psi_{i+1}}$}%
}}}}
\put(-524,-2461){\makebox(0,0)[lb]{\smash{{\SetFigFont{12}{14.4}{\familydefault}{\mddefault}{\updefault}{\color[rgb]{0,0,0}$\ket{\psi_{i+1}}$}%
}}}}
{\color[rgb]{0,0,0}\put(1426,-811){\circle*{76}}
}%
\end{picture}%
  \end{center}
  \caption{Testing that the output of $U_i$ is close to the input of
  $U_{i+1}$.  The inputs $\ket{\psi_j}$ are the ideal inputs to $U_j$,
  and are labelled for clarity only -- no assumptions are made about
  these states.
  Qubits that do not reach the right edge are traced out.}\label{fig:test}
\end{figure}
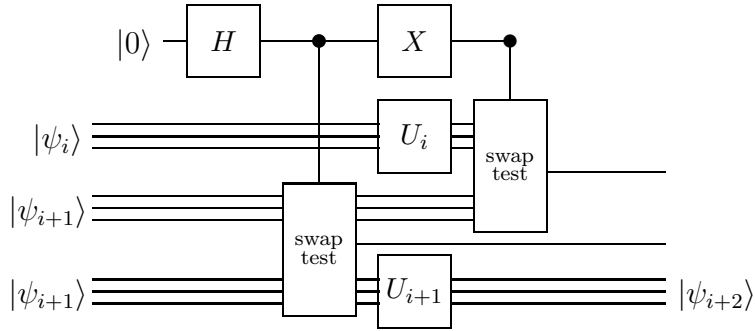

Using this decomposition of $Q_1$ and $Q_2$, circuits
$C_1$ and $C_2$ are constructed that are logarithmic in depth 
and still in some sense
faithfully implement $Q_1$ and $Q_2$.  This is done by running the
circuits corresponding to $U_1, \ldots, U_n$ in parallel, and tracing
out all the qubits that are not in the output of $U_n$.  Such a
circuit is constant depth, but does not necessarily output a state in
the image of $Q_1$, as the input to $U_i$ is not necessarily close to the
output from $U_{i-1}$.  This problem can be dealt with by comparing the
output of $U_{i-1}$ to the input to $U_{i}$.  In order to do this
in logarithmic depth an auxiliary input that is
first compared against the input to $U_i$ and then held in reserve to
compare to the output of $U_{i-1}$ is needed.  To compare these quantum states
the swap test can be used.  This test will fail with some probability depending
on the distance between the two states.
An example of the construction used to ensure that the output of
$U_{i-1}$ agrees with the input to $U_i$ is given in Figure~\ref{fig:test}.
To simplify the analysis of the constructed circuits these tests are controlled
so that either one or the other is performed.  This
will affect the failure probability by a factor of at most two, but
will allow the analysis of each swap test to ignore the effect of the
other.
To implement this a control
qubit is used so that either the first or the second test is performed
between every two pieces $U_i, U_{i+1}$ of the circuit.  
If a test is not performed, then the value of the output qubit of the
swap test is left unchanged, and so the result of the test is a qubit
in the $\ket 0$ state.
These controlled operations can be implemented in logarithmic depth using the
technique of Moore and Nilsson~\cite{MooreN02}.

\begin{figure}
  \begin{center}
\setlength{\unitlength}{3947sp}%
\begingroup\makeatletter\ifx\SetFigFont\undefined%
\gdef\SetFigFont#1#2#3#4#5{%
  \reset@font\fontsize{#1}{#2pt}%
  \fontfamily{#3}\fontseries{#4}\fontshape{#5}%
  \selectfont}%
\fi\endgroup%
\begin{picture}(5561,1824)(1114,-2473)
\put(2551,-2386){\makebox(0,0)[lb]{\smash{{\SetFigFont{12}{14.4}{\familydefault}{\mddefault}{\updefault}{\color[rgb]{0,0,0}Output of $U_n$}%
}}}}
\thinlines
{\color[rgb]{0,0,0}\put(4051,-811){\line( 1, 0){375}}
}%
{\color[rgb]{0,0,0}\put(4051,-886){\line( 1, 0){375}}
}%
{\color[rgb]{0,0,0}\put(4051,-1036){\line( 1, 0){375}}
}%
{\color[rgb]{0,0,0}\put(4051,-1111){\line( 1, 0){375}}
}%
{\color[rgb]{0,0,0}\put(4051,-1186){\line( 1, 0){375}}
}%
{\color[rgb]{0,0,0}\put(4051,-1336){\line( 1, 0){375}}
}%
{\color[rgb]{0,0,0}\put(4051,-1411){\line( 1, 0){375}}
}%
{\color[rgb]{0,0,0}\put(4051,-1486){\line( 1, 0){375}}
}%
{\color[rgb]{0,0,0}\put(4051,-2236){\line( 1, 0){375}}
}%
{\color[rgb]{0,0,0}\put(4051,-2311){\line( 1, 0){375}}
}%
{\color[rgb]{0,0,0}\put(4051,-2386){\line( 1, 0){375}}
}%
{\color[rgb]{0,0,0}\put(5026,-736){\line( 1, 0){375}}
}%
{\color[rgb]{0,0,0}\put(5026,-811){\line( 1, 0){375}}
}%
{\color[rgb]{0,0,0}\put(5026,-886){\line( 1, 0){375}}
}%
{\color[rgb]{0,0,0}\put(5026,-1111){\line( 1, 0){375}}
}%
{\color[rgb]{0,0,0}\put(5026,-1186){\line( 1, 0){375}}
}%
{\color[rgb]{0,0,0}\put(5026,-1261){\line( 1, 0){375}}
}%
{\color[rgb]{0,0,0}\put(5026,-1486){\line( 1, 0){375}}
}%
{\color[rgb]{0,0,0}\put(5026,-1561){\line( 1, 0){375}}
}%
{\color[rgb]{0,0,0}\put(5026,-1636){\line( 1, 0){375}}
}%
{\color[rgb]{0,0,0}\put(5026,-1861){\line( 1, 0){375}}
}%
{\color[rgb]{0,0,0}\put(5026,-1936){\line( 1, 0){375}}
}%
{\color[rgb]{0,0,0}\put(5026,-2011){\line( 1, 0){375}}
}%
{\color[rgb]{0,0,0}\put(5026,-2236){\line( 1, 0){375}}
}%
{\color[rgb]{0,0,0}\put(5026,-2311){\line( 1, 0){375}}
}%
{\color[rgb]{0,0,0}\put(5026,-2386){\line( 1, 0){375}}
}%
{\color[rgb]{0,0,0}\put(1126,-736){\line( 1, 0){375}}
}%
{\color[rgb]{0,0,0}\put(1126,-811){\line( 1, 0){375}}
}%
{\color[rgb]{0,0,0}\put(1126,-886){\line( 1, 0){375}}
}%
{\color[rgb]{0,0,0}\put(1126,-1036){\line( 1, 0){375}}
}%
{\color[rgb]{0,0,0}\put(1126,-1111){\line( 1, 0){375}}
}%
{\color[rgb]{0,0,0}\put(1126,-1186){\line( 1, 0){375}}
}%
{\color[rgb]{0,0,0}\put(1126,-1336){\line( 1, 0){375}}
}%
{\color[rgb]{0,0,0}\put(1126,-1411){\line( 1, 0){375}}
}%
{\color[rgb]{0,0,0}\put(1126,-1486){\line( 1, 0){375}}
}%
{\color[rgb]{0,0,0}\put(1126,-2236){\line( 1, 0){375}}
}%
{\color[rgb]{0,0,0}\put(1126,-2311){\line( 1, 0){375}}
}%
{\color[rgb]{0,0,0}\put(1126,-2386){\line( 1, 0){375}}
}%
{\color[rgb]{0,0,0}\put(2101,-736){\line( 1, 0){375}}
}%
{\color[rgb]{0,0,0}\put(2101,-811){\line( 1, 0){375}}
}%
{\color[rgb]{0,0,0}\put(2101,-886){\line( 1, 0){375}}
}%
{\color[rgb]{0,0,0}\put(2101,-1111){\line( 1, 0){375}}
}%
{\color[rgb]{0,0,0}\put(2101,-1186){\line( 1, 0){375}}
}%
{\color[rgb]{0,0,0}\put(2101,-1261){\line( 1, 0){375}}
}%
{\color[rgb]{0,0,0}\put(2101,-1486){\line( 1, 0){375}}
}%
{\color[rgb]{0,0,0}\put(2101,-1561){\line( 1, 0){375}}
}%
{\color[rgb]{0,0,0}\put(2101,-1636){\line( 1, 0){375}}
}%
{\color[rgb]{0,0,0}\put(2101,-1861){\line( 1, 0){375}}
}%
{\color[rgb]{0,0,0}\put(2101,-1936){\line( 1, 0){375}}
}%
{\color[rgb]{0,0,0}\put(2101,-2011){\line( 1, 0){375}}
}%
{\color[rgb]{0,0,0}\put(2101,-2236){\line( 1, 0){375}}
}%
{\color[rgb]{0,0,0}\put(2101,-2311){\line( 1, 0){375}}
}%
{\color[rgb]{0,0,0}\put(2101,-2386){\line( 1, 0){375}}
}%
{\color[rgb]{0,0,0}\put(4426,-2461){\framebox(600,1800){$C_2$}}
}%
{\color[rgb]{0,0,0}\put(1501,-2461){\framebox(600,1800){$C_1$}}
}%
\put(4201,-1936){\makebox(0,0)[lb]{\smash{{\SetFigFont{12}{14.4}{\familydefault}{\mddefault}{\updefault}{\color[rgb]{0,0,0}$\vdots$}%
}}}}
\put(5176,-1074){\makebox(0,0)[lb]{\smash{{\SetFigFont{12}{14.4}{\familydefault}{\mddefault}{\updefault}{\color[rgb]{0,0,0}$\vdots$}%
}}}}
\put(5176,-1824){\makebox(0,0)[lb]{\smash{{\SetFigFont{12}{14.4}{\familydefault}{\mddefault}{\updefault}{\color[rgb]{0,0,0}$\vdots$}%
}}}}
\put(1276,-1936){\makebox(0,0)[lb]{\smash{{\SetFigFont{12}{14.4}{\familydefault}{\mddefault}{\updefault}{\color[rgb]{0,0,0}$\vdots$}%
}}}}
\put(2251,-1074){\makebox(0,0)[lb]{\smash{{\SetFigFont{12}{14.4}{\familydefault}{\mddefault}{\updefault}{\color[rgb]{0,0,0}$\vdots$}%
}}}}
\put(2251,-1824){\makebox(0,0)[lb]{\smash{{\SetFigFont{12}{14.4}{\familydefault}{\mddefault}{\updefault}{\color[rgb]{0,0,0}$\vdots$}%
}}}}
\put(2551,-1036){\makebox(0,0)[lb]{\smash{{\SetFigFont{12}{14.4}{\familydefault}{\mddefault}{\updefault}{\color[rgb]{0,0,0}$\Bigg\} {\ket 0}^{\tensor n}$}%
}}}}
\put(2551,-1786){\makebox(0,0)[lb]{\smash{{\SetFigFont{12}{14.4}{\familydefault}{\mddefault}{\updefault}{\color[rgb]{0,0,0}$\Bigg\}$ Swap tests}%
}}}}
\put(5476,-1036){\makebox(0,0)[lb]{\smash{{\SetFigFont{12}{14.4}{\familydefault}{\mddefault}{\updefault}{\color[rgb]{0,0,0}$\Bigg\}$ Swap tests}%
}}}}
\put(5476,-1786){\makebox(0,0)[lb]{\smash{{\SetFigFont{12}{14.4}{\familydefault}{\mddefault}{\updefault}{\color[rgb]{0,0,0}$\Bigg\} {\ket 0}^{\tensor n}$}%
}}}}
\put(5476,-2386){\makebox(0,0)[lb]{\smash{{\SetFigFont{12}{14.4}{\familydefault}{\mddefault}{\updefault}{\color[rgb]{0,0,0}Output of $V_n$}%
}}}}
{\color[rgb]{0,0,0}\put(4051,-736){\line( 1, 0){375}}
}%
\end{picture}%
  \end{center}
  \caption{The outputs of $C_1$ and $C_2$.}\label{fig:dummy}
\end{figure}
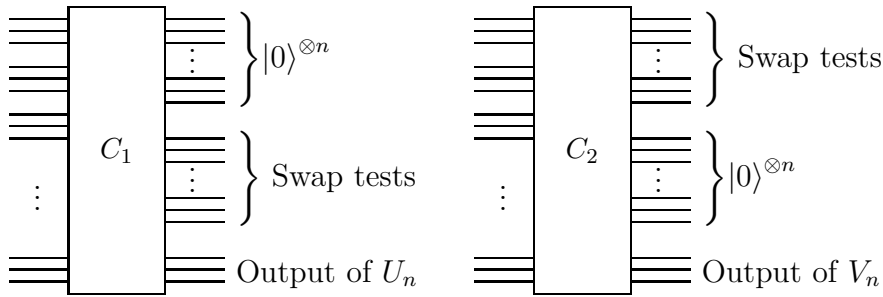

After adding these tests between each piece of the circuit there is
one final modification required.  If any of the swap tests fail, i.e. detect
states that are not the same, then they
will output qubits in the $\ket 1$ state.  As yes instances of
$\prb{CI}_{1,b}$ have outputs that are close together,
we can ensure that
no outputs of the constructed circuits can be close if any swap
tests fail by adding dummy qubits in the $\ket 0$ state to be compared
to the outputs of the swap tests in the other circuit.  These dummy
qubits are shown in Figure~\ref{fig:dummy}.

%% The circuit $C_1$ constructed from $Q_1$, including these dummy qubits, 
%% is shown in Figure~\ref{fig:logdepth}.
%% \begin{figure}
%%   \begin{center}
%%     \input{figures/logdepth.latex}
%%   \end{center}
%%   \caption{The constructed circuits $C_1$.  In the circuit $C_2$ the
%%     zero output qubits are swapped with the qubits containing the
%%     results of the swap tests.  All qubits that do not reach the right
%%     edge of the figure are traced out.}\label{fig:logdepth}
%% \end{figure}

The constructed circuits $C_1$ and $C_2$ are obtained by decomposing
$Q_1$ and $Q_2$ into constant depth pieces, inserting the swap tests shown in
Figure~\ref{fig:test}, and adding dummy qubits to ensure that the
swap tests in the other circuit do not fail.
At the end of these circuits, all qubits are traced out, except the
output (in the space $\mathcal{K}_n$) of $U_n$ or $V_n$, the output of
the swap tests, and the dummy zero qubits.  If the outputs of $C_1$
and $C_2$ are close together, then intuitively the output of the swap
tests in each circuit must be close to zero and the output of $U_n$
and $V_n$ must also be close.  If the swap tests do not fail with high
probability (i.e. the outputs are close to zero), then these circuits
will more or less faithfully reproduce the output of $Q_1$ and $Q_2$.
Thus, in the case that the outputs of $C_1$ and $C_2$ can be made close, we
will be able to argue that the output of $Q_1$ and $Q_2$ can also be
made close.  Proving that this intuitive picture is accurate forms the
content of the next section.

In the other direction, 
it is not hard to see that if there are states $\rho, \xi \in
\density{H}$ such that $Q_1(\rho) = Q_2(\xi)$, then there are similar
states for the constructed circuits $C_1$ and $C_2$.  To do this,
notice that the circuit construction does not change if additional
qubits are added to the circuits
to allow purification of the states $\rho$ and $\xi$ to be used as
inputs to $C_1$ and $C_2$.  These additional qubits are traced out
with the other qubits at the end of the circuit, so that the output
state of the circuit are not changed.
As these purifications are pure states and all operations performed
during the circuit are unitary,
the intermediate states of the circuits must also be pure states.
If the input state to $C_1$ is $\ket\psi$, then by providing the state
\[  \ket\psi \tensor U_1 \ket\psi \tensor \cdots \tensor
       U_{n-1} U_{n-2} \cdots U_1 \ket\psi \]
as input to $C_1$, the output of each block of the circuit will be
identical to the input to the next block, ensuring that all the swap
tests will succeed with probability one.
It remains only to check on such input states that $C_1$ produces the
same output as $Q_1$ on $\rho$.  This can be observed by noting that
the output of the circuit is exactly
\[ \tr_{\mathcal{B}_n} U_n U_{n-1} \cdots U_1 \rho U_1^* U_2^* \cdots U_n^*, \]
which by construction is equal to the output of $Q_1$ on $\rho$.
Thus if the circuits $Q_1$ and $Q_2$ have intersecting images then so
do the circuits $C_1$ and $C_2$.  This observation proves the
completeness of the construction.  Soundness is considerably more
intricate, and is the focus of the next section.

%=============================================================================%

\section{Soundness of the Construction}\label{scn:soundness}

In this section it is demonstrated that if the images of the original circuits
$Q_1$ and $Q_2$ are far apart then so must be the images of the
constructed circuits $C_1$ and $C_2$.  As the constructed circuits
essentially simulate $Q_1$ and $Q_2$ the desired result can be
obtained by arguing that either the outputs of $C_1$ and $C_2$ 
are far apart or
the input to at least one of the constructed circuits is not a
faithful simulation of the corresponding original circuit.  In the
case that this simulation is not faithful it will be shown that there is
some swap test that fails with reasonable probability.  This implies
that outputs of the constructed circuits must also be distant, as the
failing swap test produces a state of the form $(1-p) \ket 0 \bra 0 +
p \ket 1 \bra 1$ that has low fidelity with the
corresponding dummy zero qubit of the other circuit.

As a first step, we place a lower bound
on the failure probability of a swap test in terms of the fidelity of
the two states being compared.  In the following lemma the swap test
is viewed as a measurement of the symmetric and antisymmetric
projectors, with the outcome that produces a qubit in the state $\ket
1$ corresponding to the antisymmetric case.

\begin{lemma}\label{lem:swap-fidelity}
  If $\rho \in \density{A \tensor B}$ then a swap test
  on $\mathcal{A \tensor B}$ returns the antisymmetric outcome
  with probability at least
  \[ \frac{1}{2} - \frac{1}{2}\F(\ptr{A} \rho, \ptr{B} \rho). \]

  \begin{proof}
    Let $\ket \psi \in \mathcal{A \tensor B \tensor C}$ be a
    purification of $\rho$, where $\mathcal{C}$ is an arbitrary space
    of sufficient dimension to allow such a purification.
    The swap test measures the state on $\mathcal{A \tensor B}$ with
    the projectors $\frac{1}{2}(I - W)$ and $\frac{1}{2}(I + W)$,
    where $W$ is the swap operator on $\mathcal{A \tensor B}$.
    Thus, the antisymmetric outcome occurs with probability
    \[
      \frac{1}{4} \tr \left(
        [(I - W)\tensor I] \ket\psi\bra\psi [(I - W^*)\tensor I]
        \right)
      =  \frac{1}{2} \bra\psi I \tensor I
                      - W \tensor I \ket\psi
      =  \frac{1}{2} - \frac{1}{2}
        \bra\psi W \tensor I \ket\psi, \]
    as $W$ is hermitian.  Then as $W$ is also unitary, the states $\ket\psi$
    and $W \ket\psi$ each purify both $\ptr{A \tensor C}
    \ket\psi\bra\psi$ and $\ptr{B \tensor C} \ket\psi\bra\psi$,
    and so by Uhlmann's theorem
    \begin{align*}
      \frac{1}{2} - \frac{1}{2} \bra\psi W \tensor I \ket\psi
      & \geq \frac{1}{2} - \frac{1}{2} \F(\ptr{A \tensor C}
        \ket\psi\bra\psi, \ptr{B \tensor C} \ket\psi\bra\psi).
    \end{align*}
    After tracing out the space $\mathcal{C}$, this is exactly the
    statement of the lemma.
  \end{proof}
\end{lemma}

This lemma cannot be immediately applied to the circuits $C_1$
and $C_2$, as in these circuits the output of one block of the circuit
is not directly
compared to the input to the next block, but instead each of these
states are with probability $1/2$ compared to some intermediate value.  
In order to deal with this difficulty, we use the Fuchs-van de Graaf
inequalities to translate the fidelity to a relation involving the
trace norm, which we can then apply the triangle inequality to.  This
application of the triangle inequality shows that at least one of the
two swap tests fails with probability bounded below by an expression
involving the fidelity.  In the following corollary the reduced states
of various parts of the input to either of the circuits
$C_1$ or $C_2$ are used, but it is not assumed that these states
are given in a separable form.  For instance, the density matrices
$\rho_i, \sigma_i,$ and $\xi_i$ that appear in the lemma may be part
of some larger entangled pure state, so that the failure probabilities
of the two swap tests need not be independent.

\begin{cor}\label{cor:failprob}
  If $\ket\psi$ is input to the circuit $C_a$ for $a \in \{1,2\}$, 
  with $\rho_i$ the reduced state of $\ket\psi\bra\psi$ on
  $\mathcal{H}_i \tensor \mathcal{A}_i$, then
  at least one of the swap tests on the $i$th block
  of $C_a$ fails with probability at least
  \[ \frac{1}{64} \tnorm{U_i \rho_{i-1} U_i^* - \rho_i}^2. \]

  \begin{proof}
    In the $i$th block of $C_a$ there are two inputs to the first swap
    test: let the reduced density operators of these inputs be
    $\rho_i$ and $\sigma_i$.
    The inputs to the second swap test are then given by $\sigma_i$
    and $U_i \rho_{i-1} U_i^* = \xi_i$.
    As exactly one of these tests is performed we do not need to
    consider the effect of the first test on the state when
    considering the second test, and so the same input state
    $\sigma_i$ is used in both swap tests.

    By Lemma~\ref{lem:swap-fidelity}, the failure probability of first
    and second tests, when performed, are at least
    $\frac{1}{2}(1 - \F(\rho_i,\sigma_i))$ and
    $\frac{1}{2}(1 -\F(\sigma_i,\xi_i))$, respectively.  
    Thus, the probability $p$ that at least one of these tests
    fails, given that each of them is performed with probability
    $1/2$, is at least
    \[p \geq \frac{1}{2} \max \left\{ \frac{1}{2}(1 -\F(\sigma_i,\xi_i)),
           \frac{1}{2}(1 -\F(\rho_i, \sigma_i)) \right\}
      = \frac{1}{4} \left(1 - \min\{\F(\sigma_i, \xi_i)),
           \F(\rho_i, \sigma_i)\}\right). \]
    By the Fuchs-van de Graaf inequalities, this fidelity may be
    replaced by the trace norm.  Doing so, we obtain
    \[ p \geq \frac{1}{16} \max( \tnorm{\sigma_i - \xi_i}^2,
           \tnorm{\rho_i - \sigma_i}^2 ). \]
    Finally, as this maximum must be at least the average of the two
    values,
    \[ p \geq \frac{1}{16} \left(\frac{\tnorm{\sigma_i - \xi_i}}{2}
           + \frac{\tnorm{\rho_i - \sigma_i}}{2} \right)^2
       \geq \frac{1}{64} \tnorm{\rho_i - \xi_i}^2, \]
    where the last inequality follows from an application of the
    triangle inequality.
  \end{proof}
\end{cor}

By repeatedly applying some of the properties of the trace norm
discussed in Section~\ref{scn:prelim} it is somewhat tedious but not
difficult to reduce the problem at hand to the previous Corollary.
This is the content of the following theorem.

\begin{theorem}\label{thm:soundness}
  If $\F(Q_1(\rho_0), Q_2(\xi_0)) < 1 - c$ for all $\rho_0, \xi_0 \in \mathcal{H}$ then 
  \[ \F(C_1(\rho), C_2(\xi)) <  1 - \frac{c^2}{144 n^2} \]
  for all $\rho, \xi \in (\mathcal{H \tensor A})^{\tensor 2n}$.

  \begin{proof}
    Let $\rho$ and $\xi$ be inputs to $C_1$ and $C_2$, and let
    $\rho_i, \xi_i$ be the reduced states of these inputs on 
    $\mathcal{H}_i \tensor \mathcal{A}_i$ for $0
    \leq i \leq 2n$, where the states for $i > n$ are the
    inputs that are only used by the swap tests, which we will not
    need to refer to explicitly.
    That is, $\rho_i$ and $\xi_i$ for $0 \le i \le n$ 
    are the portions of the state that are
    input to the unitaries $U_i$ and $V_i$ that make up the circuits
    $Q_1$ and $Q_2$.
    The output of the circuits $C_1$ and $C_2$ is
    then given by a number of qubits corresponding to the swap tests
    as well as the states $\tr_{\mathcal{B}_n}
    \rho_n$ and $\tr_{\mathcal{B}_n} \xi_n$, where $\mathcal{B}_n$ is
    simply the space that is traced out to obtain the output from the
    unitary representations of the original circuits.

    By the condition on the fidelity of $Q_1$ and $Q_2$ and the
    Fuchs-van de Graaf inequalities, we have
    $ 2c < \tnorm{ Q_1(\rho_0) - Q_2(\xi_0) }. $
    Using the triangle inequality we can relate this to the
    distance between the constructed circuits.  Adding terms and
    simplifying, we obtain
    \begin{align*}
      2c &< \tnorm{ Q_1(\rho_0) 
	           -\tr_{\mathcal{B}_n}\rho_n 
		   +\tr_{\mathcal{B}_n}\xi_n 
		   - Q_2(\xi_0)
	           +\tr_{\mathcal{B}_n}\rho_n 
		   -\tr_{\mathcal{B}_n}\xi_n } \\
        &\le \tnorm{ Q_1(\rho_0)
	           -\tr_{\mathcal{B}_n}\rho_n }
	  +\tnorm{ \tr_{\mathcal{B}_n}\xi_n 
		   - Q_2(\xi_0)}
	  +\tnorm{ \tr_{\mathcal{B}_n}\rho_n 
		   -\tr_{\mathcal{B}_n}\xi_n } .
    \end{align*}
    We now observe that $\tnorm{ \tr_{\mathcal{B}_n}\rho_n
    -\tr_{\mathcal{B}_n}\xi_n } \leq \tnorm{ C_1(\rho) - C_2(\xi) }$
    by the monotonicity of the trace norm under the partial trace,
    since the former can be obtained from the later by tracing out the
    appropriate spaces.  Using this we have
    \begin{align}
    2c < \tnorm{ Q_1(\rho_0)
	           -\tr_{\mathcal{B}_n}\rho_n }
	  +\tnorm{ \tr_{\mathcal{B}_n}\xi_n 
		   - Q_2(\xi_0)}
	  +\tnorm{ C_1(\rho) - C_2(\xi) } \label{eqn:three-terms-2c}
    \end{align}
    As the three terms on the right are nonnegative, at least one of them
    must be larger than the average $2c/3$.  
    If $\tnorm{ C_1(\rho) - C_2(\xi) } > 2c/3$
    then $\F(C_1(\rho), C_2(\xi)) < 1 - c^2/144$ and there
    is nothing left to prove.

    The cases where one of the first two terms of
    \eqref{eqn:three-terms-2c} exceeds $2c/3$
    are symmetric, and so we can consider only the first term.
    Expanding $Q_1(\rho_0)$ in terms of the $U_i$, we obtain
    \begin{align*}
      \frac{2c}{3} 
      &< \tnorm{ Q_1(\rho_0) - \tr_{\mathcal{B}_n}\rho_n }\\
      &= \tnorm{ 
	    \tr_{\mathcal{B}_n} 
          U_n U_{n-1} \cdots U_1 \rho_0 U_1^* U_2^* \cdots U_n^*
 	   -\tr_{\mathcal{B}_n} \rho_n } \\
      &\le \tnorm{ U_n U_{n-1} \cdots U_1 \rho_0 U_1^* U_2^* \cdots U_n^*
	    -\rho_n },
    \end{align*}
    where once again the monotonicity of the trace norm
    under the partial trace has been used.
    By repeating the strategy of adding terms and then applying the
    triangle inequality we have
    \begin{align*}
      \frac{2c}{3}
      &< \tnorm{U_1 \rho_0 U_1^* - \rho_1} + \tnorm{
	    U_{n} U_{n-1} \cdots U_2 \rho_1 U_2^* U_3^* \cdots U_{n}^*
	    -\rho_n}.
    \end{align*}
    Here we have made use of the unitary invariance of the trace norm
    to discard the operators $U_2, \ldots U_n$ from the first term.
    Continuing in this fashion we have
    \begin{align*}
      \frac{2c}{3}
      &< \sum_{i=1}^n \tnorm{U_i \rho_{i-1} U_i^* - \rho_i}.
    \end{align*}
    As all terms in this sum are nonnegative,
    there must be at least one term in the sum that exceeds
    $2c / (3n)$, as this is a lower bound on the average of all terms.
    Thus, for some value of $i$, we have
    $ \tnorm{U_i \rho_{i-1} U_i^* - \rho_i} > 2c / (3n), $
    and so by Corollary~\ref{cor:failprob} one of the
    corresponding swap tests fails with probability
    $p > c^2 / (144 n^2)$.  The qubit representing the output value of
    this swap test is then of the form $(1-p) \ket 0 \bra 0 + p \ket 1
    \bra 1$, and so, by the monotonicity of the fidelity under
    the partial trace,
    \[ \F(C_1(\rho), C_2(\xi)) 
       \leq \F((1-p) \ket 0 \bra 0 + p \ket 1 \bra 1, \ket 0 \bra 0)
       = 1-p
       < 1 - \frac{c^2}{144 n^2}, \]
    as in the statement of the theorem.
  \end{proof}
\end{theorem}

By combining Theorem~\ref{thm:soundness} with the observation in
Section~\ref{scn:construction} and the
multiplicativity of the maximum output
fidelity of two transformations, we obtain the following result.

\begin{cor}
  The problem $\prb{Log-depth CI}_{1,b}$ is \class{QIP}-complete
  for any constant $0 < b < 1$.

  \begin{proof}
    Theorem~\ref{thm:soundness} establishes the completeness of the
    problem for any $b \geq 1 - c^2/(144 n^2)$, where $n$ is an upper
    bound on the size of the circuits.  Using
    Theorem~\ref{thm:outfid-mult} of Kitaev, Shen, and
    Vyalyi~\cite{KitaevS+02} we can repeat each of the circuits $r$
    times in parallel to obtain the completeness of the problem for $b
    \geq \left( 1 - c^2/(144 n^2) \right)^r, $ which can be made
    smaller than any constant for $r$ some polynomial in $n$.
  \end{proof}
\end{cor}

As the circuits constructed by the reduction only make use of
logarithmic depth when performing swap tests, and the controlled swap
operations performed by these tests can be accomplished in constant
depth using unbounded fan-out gates, the following Corollary follows
immediately from the previous one.

\begin{cor}
  The problem $\prb{Const-depth CI}_{1,b}$ on circuits with the
  unbounded fan-out gate is \class{QIP}-complete for
  for any constant $0 < b < 1$.
\end{cor}

%=============================================================================%

\section{Distinguishing Log-Depth Computations}\label{scn:distinguish}

The hardness of $\prb{Log-depth CI}_{1,b}$ can be extended to
$\prb{Log-depth QCD}_{2,b}$ by observing that the reduction for the
polynomial depth version of this problem in~\cite{RosgenW05} can be
made to preserve the depth of the constructed circuits.  Once this
observation is made, the hardness of the log-depth (and constant-depth
with fan-out) versions of the circuit distinguishability problem is
immediate.

The reduction in~\cite{RosgenW05} takes as input circuits $(Q_1, Q_2)$
and produces circuits $C_1$ and $C_2$.  Without describing the
reduction in detail, the constructed circuits $C_1$ and $C_2$ run,
depending on the value of a control qubit, one of $Q_1$ and $Q_2$,
followed by a constant depth circuit.
If the input circuits $Q_1$ and $Q_2$ have logarithmic depth, then the
only significant difficulty is the fact that controlled versions of
these circuits are needed.
However, as we have already seen, if we replace the gates in $Q_1$ and
$Q_2$ with controlled versions, then we can use the scheme of Moore
and Nilsson~\cite{MooreN02} to implement the controlled operations in
logarithmic depth.
With this modification, the reduction in~\cite{RosgenW05} can be
reused to show the hardness of the \prb{QCD} problem on log-depth
circuits.

\begin{cor}
  $\prb{Log-depth QCD}_{2,b}$ is \class{QIP}-complete
  for any constant $0 < b < 2$.
\end{cor}

Once again these controlled operations can be implemented in a constant
depth circuit if the unbounded fan-out gate is allowed into the set of allowed
gates.

\begin{cor}
  $\prb{Const-depth QCD}_{2,b}$ on circuits with the
  unbounded fan-out gate is \class{QIP}-complete
  for any constant $0 < b < 2$.
\end{cor}

%=============================================================================%

\section{Conclusion}\label{scn:concl}

The hardness of distinguishing even log-depth mixed state quantum
circuits leaves several related open problems, a few of which are listed here.
\begin{itemize}
  \item Can this new complete problem be used to further understand
    \class{QIP}?
  \item Does this result rely in an essential way on the mixed state
     circuit model?  How difficult is it to distinguish quantum
     circuits in less general models of computation?
  \item What is the complexity of distinguishing constant depth
     quantum circuits that do not use the unbounded fan-out gate?   
\end{itemize}

%=============================================================================%

\section*{Acknowledgement}
I would like to thank John Watrous for several helpful discussions,
the anonymous reviewers for constructive comments,
as well as Canada's NSERC and MITACS for supporting this
research.

\end{document}